# Spontaneous emission of radiation by metallic electrons in the presence of electromagnetic fields of surface plasmon oscillations

Sándor Varró[*], Norbert Kroó, Győző Farkas and Péter Dombi

*Research Institute for Solid State Physics and Optics of the Hungarian Academy of Sciences*

*Budapest, Hungary*

The spontaneous emission of metallic electrons embedded in a high-intensity enhanced surface plasmon field is considered analytically. The electrons are described by dressed quantum states which contain the interaction with the plasmon field non-perturbatively. Considerable deviations from the perturbative behaviour have been found in the intensity dependence of the emitted fundamental and the second harmonic signals, even at moderate incoming laser intensities. The theoretical predictions deduced from the formalism are in good qualitative agreement with the experimental results.

**Keywords:** field enhancement, surface nonlinear optics, surface plasmons

[*] Corresponding author. Email: varro@mail.kfki.hu

## 1. Introduction

Surface electromagnetic radiation bound to a plane interface separating a dielectric and a conductor started to receive considerable attention just 100 years ago, when a paper by Sommerfeld appeared on the effect of the Earth on the propagation of radio waves [1-2][1]. In the meantime such and similar surface waves have received the name surface plasmon polaritons or surface plasmon oscillations (SPO). As is well-known, this oscillatory fields are normal modes of the electromagnetic (EM) radiation in systems consisting of dielectrics (e.g. glass) and conductors (e.g. metals or semiconductors) which are separated by sharp interfaces. Already before the sixties of the last century the number of theoretical and experimental studies of SPOs was growing fast, and nowadays the SPO physics is an important sub-discipline of electrodynamics and optics. On the early development of this branch of research an excellent collections of papers can be found e.g. in the books edited by Burstein and de Martini [3] and by Maradudin et al. [4], including many important references. For a good exposition of the subject see also the books by Raether [5]. The more recent results are summarized e.g. in the extensive and thoroughly written reports by Zayats et al. [6-7]. Surface plasmon physics still raises many fundamental questions concerning radiation-matter interaction and optics [8-16], which are interesting in themselves. At the same time, this rapidly growing field of research will surely impact many branches of science and technology [17].

One of the main characteristics of the SPOs is their very large electromagnetic fields concentrated at the interfaces of metals and dielectrics. To our knowledge, such an enhancement was first discussed by Fano [18] in 1938. Recently this phenomenon has become the subject of extensive research, because on the basis of it high-order nonlinear processes can be induced even at relatively moderate intensities of an incomig radiation which excites the SPOs [19-25]. For instance in Ref. [22] the authors report on the observation of high-harmonic generation (HHG) from an argon gas jet, in the extreme ultraviolet, induced by a field of a Ti: Sapphire laser pulse, which was resonantly enhanced through surface plasmons within a metallic nanostructure. As for the theory of such experimental findings, one would think at first glance that, at least from the point of view of the generation itself by a *single atom*, there are no new elements needed to interpret this results, because the interpretation of HHG has long received a well-established theoretical framework [26-28]. However, in the case of surface plasmon excitation the modal structure of the radiation near the surface is considerably different from that of in free space, and this circumstance has to be taken into account [29], too. The HHG process induced *directly* by high-intensity laser fields on *metal surfaces* of thick samples has also been a subject of extensive research earlier [30], and both classical and quantum mechanical [31-32] interpretations have been worked out. On the other hand, if a *thin metal layer* is excited through a dielectric (say in the Kretschmann geometry [33]), the enhanced field of the generated surface plasmons may govern the light emission process. The light emission can be quite satisfactorily interpreted e.g. as resulting due to the surface roughness, which secures the momentum conservation in the 'light to SPO' and 'SPO to light' conversion processes. In the quantum description there appears an interaction Hamiltonian due to surface roughness which already couples these two different kinds of electromagnetic radiation [34]. Usually the coupling is described by a phenomenological interaction term, and there are several techniques to generate SPOs through surface-geometrical effects, as is summarized e.g. by Zayats et al. [7]. In our present paper we study the quantum mechanical coupling of a strong external plasmon field to the photon field through an assembly of free electrons, and we do not *explicitly* take into account surface roughness. According to this description, the high-harmonic generation on thin metal films is a result of multiple-plasmon scattering on a free electron gas inside the metal. This may seem to be an analogous process to the multiphoton Compton scattering by free electrons, which has an extensive literature (see e.g. [35] and references therein). More precisely, the process to be considered is spontaneous emission of photons by an electron reflected by the metal surface in the presence of a high-intensity SPO field. At this point let us note that very likely several theoretical approaches, which have long been worked out to treat nonlinear processes taking place directly in high-intensity laser fields, could be implemented in the recent investigations of processes induced by the strong enhanced SPO fields. The present paper may be considered as a contribution belonging to this direction of the theoretical research .

---

[1] Stratton [1] writes at the beginning of Section 9.28 of his book that "The classical investigation of the effect of a finitely conducting plane upon the radiation of an oscillating dipole was published by Arnold Sommerfeld in 1909. Since that time an enormous amount of work has appeared on the subject and it may be fairly said no aspect of the problem of radio wave propagation has received more careful attention." On pages 584-585 a brief summary of the early history of electromagnetic surface waves is given, and in Section 9.28 a detailed discussion of the problem is presented. Section 11 of Sommerfeld's fundamental paper [2] is devoted to the discussion of the analogies between the propagation of *radio waves* and *optical waves* in the vicinity of a plane interface separating a dielectrics and a conducting medium (metal). This was the first hint to that surface waves may play an important role in the visible part of the electromagnetic spectrum.

In continuing our recent work on light emission by surface plasmon oscillations [36-37], the immediate motivation of the present study have been the appearently strange results of our experiments [38] in which we measured the intensity dependence of light emission (fundamental and second harmonic) generated by a moderately intense laser radiation from a thin film of gold on a glass prism. The measured slope of the signal-intensity curve manifestly deviated from that, one would expect on the basis of earlier treatments, even if one takes non-perturbatively into account the field enhancement, too. In the following we shall give a theoretical description of the light emission process, in the frame of which the basic features of the above-mentioned experimental results can be interpreted.

In Section 2 we shall briefly summarize the main steps leading to the explicit form of the electromagnetic fields of the SPOs, and give a physical background for the key parameters naturally appearing in the analysis. A particular emphasize shall be put on the clear distinction between the scattering (evanescent) waves of the EM radiation and the SPO fields generated at the metal-vacuum interface. Section 3 is devoted to the discussion of the general analytic form and physical content of the wave functions of an electron dressed by the surface plasmon field. In Section 4 the transition probabilities of the spontaneous emission of photons by an electron reflected at the metal surface in the presence of the high-intensity SPO field are given. The analytic results obtained shall be used in a few numerical examples to illustrate the intensity dependence, the degree of nonlinearity and the angular distribution of the emitted light signal. In Section 5 a brief summary closes our paper.

## 2. Generation of the enhanced scattering fields and SPOs

In the present section we give a quantitative picture of the enhanced fields appearing at plane interfaces separating a metal layer and dielectrics. The spatial distribution of the SPO fields in single- and multiple-film structures and their dispersion relations have been investigated in details e.g. by Economou for losless systems [39]. Burke et al. [40] have given a thorough analysis of propagation of SPO-like waves guided by thin, lossy metal films, and they have also discussed the question of wave launching. Here we will not enter into the discussion of this important problem, rather, we shall merely summarize the basic 'kinematic' characterization of the fields.

In the Kretschmann geometry [33] we have to consider the solutions of the Maxwell equations in three regions shown in Figure 1. Region 1 ($z > d$) is filled with a dielectric, which represents e.g. a glass prism of index of refraction $n_1 = \varepsilon_1^{1/2}$. The metal layer occupies region 2 ($0 < z < d$) of thickness $d$ and of (in general complex) dielectric constant $\varepsilon_2 = n_2^2$. In the experiments region 3 ($z < 0$) is often simply air, which can practically be considered as vacuum with the dielectric constant $\varepsilon_3 = 1$. The solutions of the Maxwell equations automatically fall into two classes of independent waves; the transverse magnetic (TM: **B** is perpendicular to the $x-z$ scattering plane) and the transverse electric (TE: **E** is perpendicular to the $x-z$ scattering plane) waves, which are also called p-polarized and s-polarized waves, respectively. In each cases both $E_x$ and $B_y$ have to be continuous at the glass-metal ($z=d$) and the metal-air ($z=0$) interfaces. If we take a common pure harmonic temporal dependence $e^{-i\omega t}$ of the TM fields, then for $B_y$ the Helmholtz equation $(\partial_x^2 + \partial_z^2 + K^2)B_y = 0$ follows, where $K^2 \equiv (\omega/c)^2 \varepsilon$. In the scattering problem under discussion the fields are usually modelled by superpositions of plane waves of the form $\exp[i(K_x x + K_z z)]$ with four unknown constant amplitudes determined by the *inhomogeneous* set of linear algebraic equations stemming from the boundary conditions. In Figure 1 the capital letters $F_1$, $G_2$, $F_2$ and $G_3$ label the reflected, the first refracted, the second reflected and the second refracted (transmitted) waves, respectively. These waves are directly induced by the incoming field of amplidude $F_0 \neq 0$, and they represent *scattering states*. If the angle of incidence $\theta_0$ is smaller than the angle of total reflection $\theta_t$ (see below), then $G_3$ represents a free running wave. For $\theta_0$ larger than $\theta_t$, this refracted wave becomes evanescent (bound to the metal-air interface). The small letters $f_1$, $g_2$, $f_2$ and $g_3$ denote the amplitudes characterizing the SPO *eigenmodes* in the dielectric-metal-vacuum system. They are solutions of a set of *homogeneous* linear algebraic equations ($F_0 = 0$) whose determinant must vanish in order to have nontrivial solutions. As is suggested in the figure, all these latter waves are evanescent.

According to the above considerations, the complex magnetic induction and the electric field strength of a p-polarized plane wave in regions 1, 2 and 3 can be explicitly expressed as

$$\mathbf{B} = \mathbf{e}_y B(z) e^{i(K_x x - \omega t)}, \quad \mathbf{E} = (c/i\omega\varepsilon)[\mathbf{e}_x \partial_z B(z) - i\mathbf{e}_z K_x B(z)] e^{i(K_x x - \omega t)}. \tag{1}$$

Here $\mathbf{e}_{x,y,z}$ denote unit vectors pointing along the corresponding positive coordinate axes. In the above equation we can *either* take $K_x = K_0 \sin\theta_0$ with $K_0 \equiv (\omega_0/c)\varepsilon_1$, *or* $K_x = k_{1x}$, depending on whether we are dealing with scattering waves or with the SPO eigenmodes of wave numbers $k_{sp} \equiv k_{1x}(\omega)$. The boundary conditions (the continuity of $E_x$ and $B_y$ at the interfaces) are equivalent to the requirement that $B(z)$ and $\partial_z B(z)$ be continuous at both of the planes defined by the relations $z = d$ and $z = 0$.

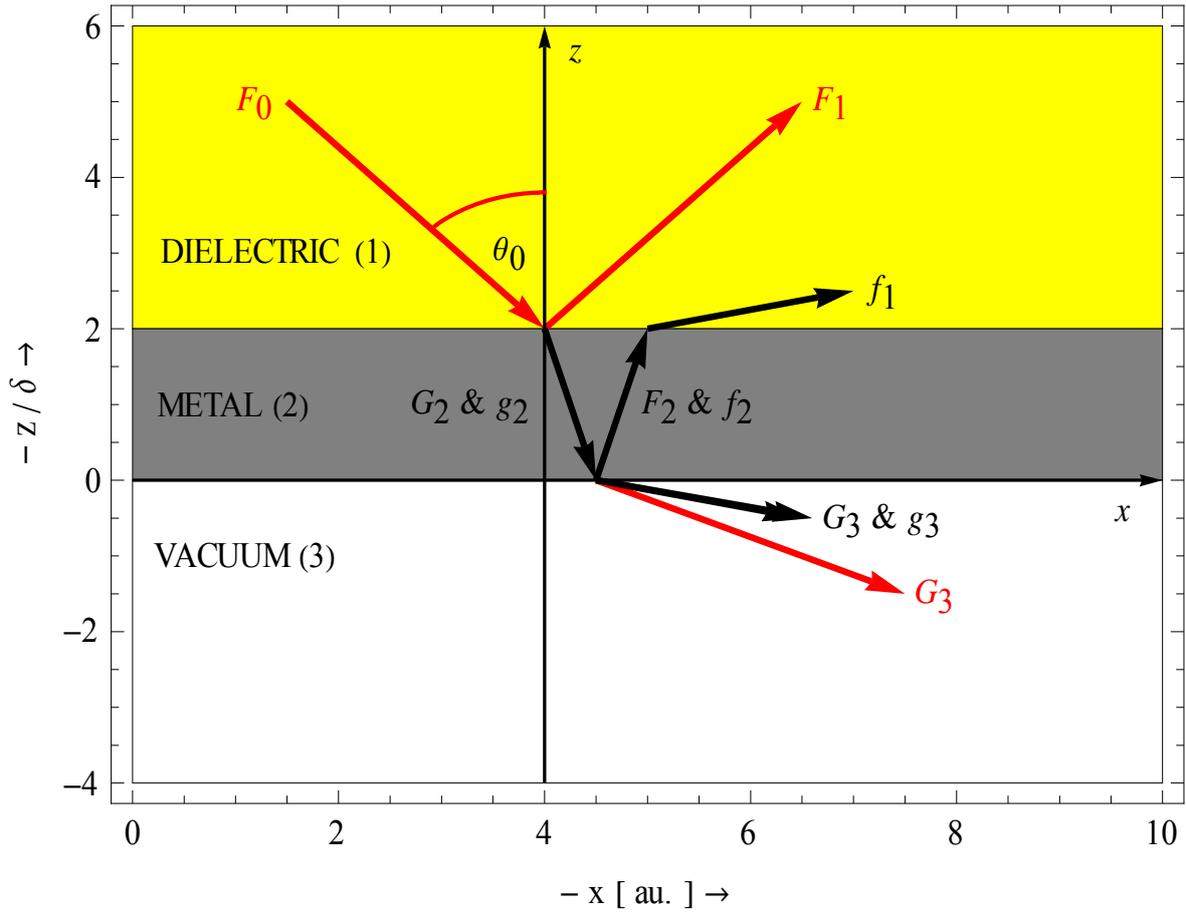

**Figure 1.** Illustrates the TM electromagnetic field configuration excited by a plane wave of amplitude $F_0$ impinging on the metal layer from the dielectric at an angle of incidence $\theta_0$. The $z$-coordinate is measured in units of the skin depth $\delta$, which is typically $22.5 nm$ in our experiments. The freely propagating waves are symbolized by red arrows, and the evanescent waves are symbolized by black arrows. For further details see the main text.

As is illustrated in Fig. 1, the scattering states are represented by the wave functions

$$B_0(z) \equiv \begin{cases} n_1(F_0 e^{-iK_{1z}z} + F_1 e^{+iK_{1z}z}), & (z > d) \\ n_2(G_2 e^{-iK_{2z}z} + F_2 e^{+iK_{2z}z}), & (0 < z < d), \\ n_3 G_3 e^{-iK_{3z}z}, & (z < 0) \end{cases} \quad \begin{aligned} K_{(1,2,3)z} &\equiv (\omega_0/c)\sqrt{\varepsilon_{(1,2,3)} - \varepsilon_1 \sin^2\theta_0} \\ &= i(\omega_0/c)\sqrt{\varepsilon_1 \sin^2\theta_0 - \varepsilon_{(1,2,3)}} \end{aligned}, \quad (2)$$

where $n_1(\omega_0/c)\sin\theta_0 = K_{1x} \equiv K_x$ is the $x$-component of the wave vector of the incoming field. If $\theta_0$ is larger than the angle of total reflection $\theta_t$ (defined by the equation $n_3 = n_1 \sin\theta_t$, and $\theta_t = 41.81°$ in our case), then $K_{3z} = +i|K_{3z}|$ is purely imaginary, and $G_3$ corresponds to an evanescent wave in region 3. The wave functions given by Eq. (2) represent the electromagnetic field configuration induced by an incoming radiation whose electric field strength reads

$$\mathbf{E_0} = (\mathbf{e_x}\cos\theta_0 + \mathbf{e_z}\sin\theta_0)|F_0|\cos\left[\omega_0\left(t - n_1\frac{x\sin\theta_0 - z\cos\theta_0}{c}\right) + \varphi_0\right], \quad (3)$$

where $\varphi_0$ is the phase of the complex amplitude $F_0$.

The wave functions $B_{sp}(z)$ of the SPO eigenmodes look similar to $B_0(z)$ of Eq. (2), but there are also crucial differences between them. On one hand, they do not contain the inhomogenity term, and, on the other hand, they are evanescent both in regions 1 and 3, i.e. they are bound to both of the interfaces, moreover, in general they spatially decay along the positive $x$-direction. The mode function of the SPO reads

$$B_{sp}(z) \equiv \begin{cases} n_1 f_1 e^{-k_{1z}z}, & (z > d) \\ n_2(g_2 e^{k_{2z}z} + f_2 e^{-k_{2z}z}), & (0 < z < d), \\ n_3 g_3 e^{k_{3z}z}, & (z < 0) \end{cases} \quad \begin{aligned} k_{(1,2,3)z} &= \sqrt{k_{1x}^2 - (\omega/c)^2 \varepsilon_{(1,2,3)}} \\ &\equiv (\omega/c)\sqrt{\varepsilon_1 \sin^2\theta' - \varepsilon_{(1,2,3)}} \end{aligned}, \quad (4)$$

with $k_{1x} \equiv k_{sp}$ and $\text{Re}[k_{(1,3)z}] > 0$. In Eq. (4) we have introduced the complex angle $\theta'$ by the definition $k_{sp} \equiv (\omega/c)n_1 \sin\theta'$ in order to emphasize the *formal* analogy between the formulas to that of Eq. (2). From the set of four linear algebraic equations, obtained from the boundary conditions, the amplitudes $F_1$, $G_2$ and $F_2$ can be expressed in terms of $G_3$, where

$$G_3 = (4n_1 B/n_3)e^{i(K_{2z}-K_{1z})d} F_0 / D_0, \quad D_0 \equiv (A+1)(B+1) - (A-1)(B-1)e^{2iK_{2z}d}, \quad (5a)$$

$$F_1 = [(A+1)(B-1) - (A-1)(B+1)e^{2iK_{2z}d}] \cdot F_0 e^{-2iK_{1z}d}/D_0, \quad F_2 = (n_3/2n_2)(1-A)G_3, \quad (5b)$$

$$G_2 = (n_3/2n_2)(1+A)G_3, \quad A \equiv (n_2 K_2/K_{2z})(K_{3z}/n_3 K_3), \quad B \equiv (n_2 K_2/K_{2z})(K_{1z}/n_1 K_1). \quad (5c)$$

Similarly, $f_1$, $g_2$ and $f_2$ can be expressed in terms of $g_3$, but now, in the homogeneous case we have an indeterminate equation for $g_3$,

$$0 = D_{sp} \cdot g_3, \quad D_{sp} \equiv (a+1)(b+1) - (a-1)(b-1)e^{-2k_{2z}d}, \quad (6a)$$

$$f_1 = [(1+a) + (1-a)e^{-2k_{2z}d}] \cdot g_3 e^{+k_{2z}d + k_{1z}d}, \quad f_2 = (n_3/2n_2)(1-a)g_3, \quad (6b)$$

$$g_2 = (n_3/2n_2)(1+a)g_3, \quad a \equiv (n_2 k_2/k_{2z})(k_{3z}/n_3 k_3), \quad b \equiv (n_2 k_2/k_{2z})(k_{1z}/n_1 k_1). \quad (6c)$$

The determinants $D_0$ and $D_{sp}$ defined in Eqs. (5a) and (6a), respectively, correspond to two distinct, mathematically and physically different situations. The scattering problem can have a physically acceptable solution only if $D_0 \neq 0$ (in the expression of $G_3$ in Eq. (5a), the denominator must not be zero). On the other hand, the homogeneous problem (corresponding to the free eigenoscillations of the dielectric-metal-dielectric system) can have nontrivial solutions only if the determinant $D_{sp} = 0$ vanishes. As is well known, this latter condition defines the propagation constant $k_{sp}(\omega)$ of the SPO through a transcendental equation. If $|k_{2z}d| \gg 1$ (i.e. in the case of a relatively thick layer), according to Eq. (6a), the condition $D_{sp} = 0$ becomes approximately the equation $a = -1$ (from which the well-known implicit dispersion relation $k_{sp} = (\omega/c)[\varepsilon_2 \varepsilon_3 /(\varepsilon_2 + \varepsilon_3)]^{1/2}$ can be derived). Quite similarly, if in Eq.(5a) $|\exp(2iK_{2z}d)| \ll 1$, then the determinant $D_0$ (and hence the denominator in the expression for $G_3$) gets close to zero if $A$ approaches $-1$. If we take $\varepsilon_2 = -\varepsilon_R + i\varepsilon_I$, with both $\varepsilon_R$ and $\varepsilon_I$ positive, and, moreover if we assume that $\varepsilon_I \ll \varepsilon_R$, then in the exponent we have approximately $-(4\pi d/\lambda_0)\sqrt{\varepsilon_R + 1}$. The factor in front of the square root is 0.711 in our case, where $d = 45 nm$, and $\lambda_0 = 795 nm$, i.e. in cases we are interested in, the exponential factor is much less than unity ($e^{-(4\pi d/\lambda_0)\sqrt{\varepsilon_R + 1}} = 0.025$ if we take $\varepsilon_R = 25.81$ for a gold layer). By varying the parameters of the incoming field the amplitudes may undergo a resonant change (e.g. enhancement) in the ranges where the determinant $D_0$ gets close to zero. The value of this determinant is governed by the parameters $A$ and $B$, whose dependence on the angle of incidence $\theta_0$ is displayed in Fig. 2 in the special case $\varepsilon_1 = (1.5)^2$, $\varepsilon_2(\omega_0) = -\varepsilon_R + i\varepsilon_I = -25.82 + i1.63$ and $\varepsilon_3 = 1$. The numerical value of $\varepsilon_2$ has been taken from Johnson and Christy [41].

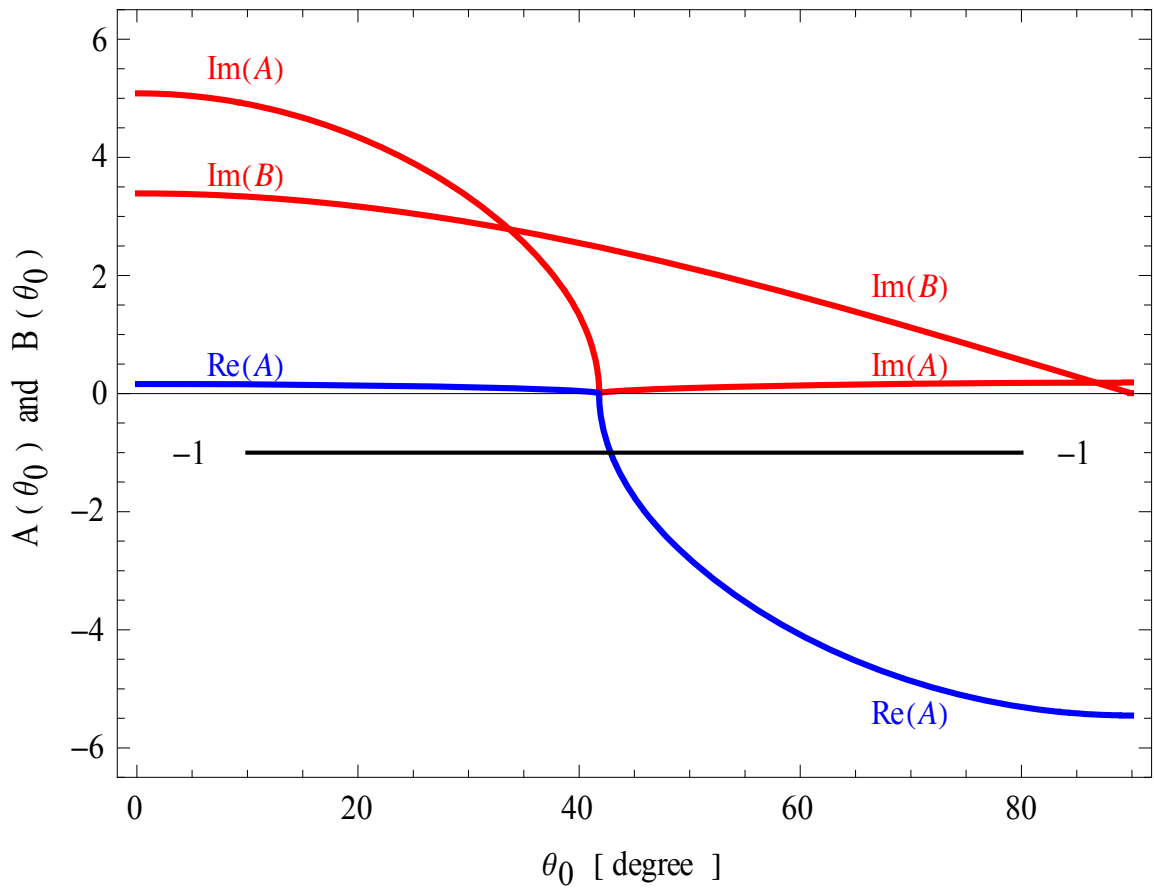

**Figure 2.** Shows the dependence on the angle of incidence $\theta_0$ of the parameters $A$ and $B$ defined in the second raw of Eq. (5). Besides the $\theta_0$-dependence, these parameters depend on the frequency of the incoming radiation through the permittivities. We have taken $\varepsilon_1 = 2.25$, $\varepsilon_2 = -25.82 + i1.63$ and $\varepsilon_3 = 1$. We note that $A$ and $B$ do not depend on the thickness $d$ of the metal layer. The dielectric in region 1 represents a glass substrate of index of refraction $n_1 = 1.5$, the metal layer in region 2 is assumed to be gold, and for region 3 we have taken vacuum. The exciting radiation field comes from a Ti:Sapphire laser of wavelength 795 nm whose photon energy is $1.5627 eV$. The curves labeled by "Im(A)", "Re(A)" and "Im(B)" represent the imaginary and the real part of $A$ and the imaginary part of $B$, respectively. The horizontal thin line labeled by "$-1$" represents the constant function $= -1$. At the angle of total reflection $\theta_t = 41.81°$ the parameter $A$ changes from almost purely imaginary to almost purely negative real numbers, and at the critical angle $\theta_c = 42.84° = \theta_t + 1.03°$ its real part takes the value $-1$. The parameter $B$ is almost purely imaginary throughout the whole angular range $0° \leq \theta_0 < 90°$, that is why we have not shown its real part.

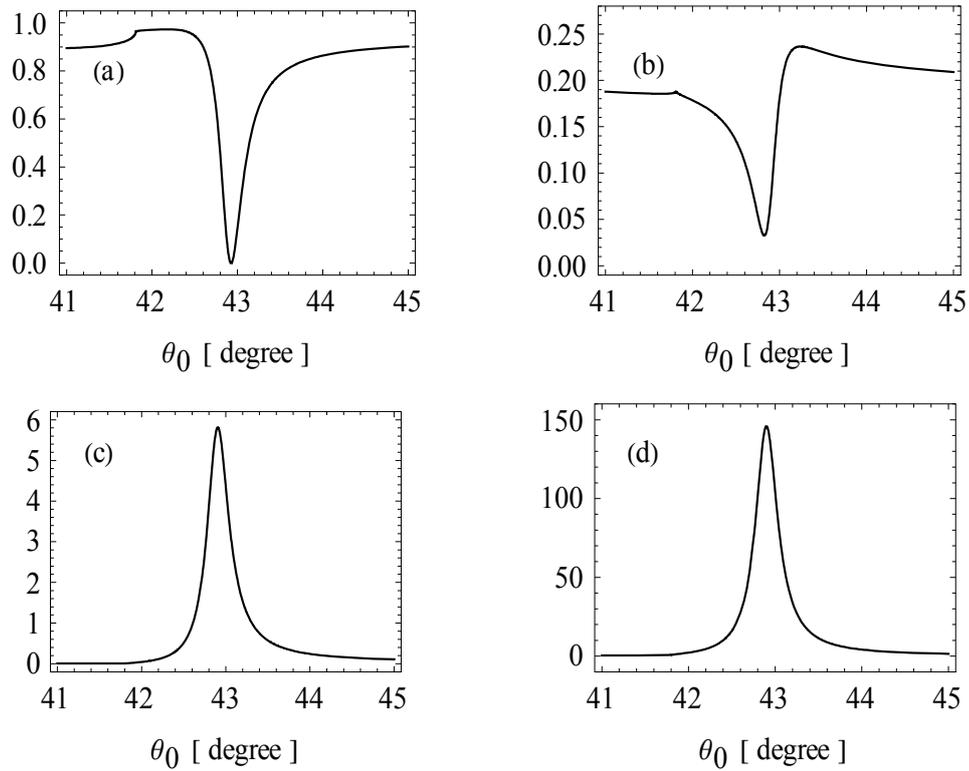

**Figure 3.** Shows the relative modulus squared of the amplitudes of the field configuration generated by an incoming plane wave as functions of the angle of incidence. (a): $|F_1/F_0|^2$, (b): $|G_2/F_0|^2$, (c): $|F_2/F_0|^2$ and (d): $|G_3/F_0|^2$ show the relative intensites of the reflected, refracted, secondary reflected and the secondary refracted (transmitted) wave, respectively. The thickness $d$ of the metal layer has been taken to be $2\times\delta=45$nm, the other parameters used in the calculations are the same as in Figure 2.

In Figure 3 we have plotted the relative modulus squared of the amplitudes of the induced field components as functions of the angle of incidence. At the critical angle $\theta_c = 42.84° = \theta_t + 1.03°$ the amplitude $F_1$ of the reflected wave drops practically to zero in a narrow angular range of half width of about $0.4°$. In Fig. 3(a) we see a clean example of the attenuated total reflection (ATR) in the Kretschmann geometry in the case of ideal plane interfaces. Fig. 3(d) shows that around $\theta_c$ the transmitted intensity $\sim |G_3|^2$ undergoes a considerable resonant enhancement of magnitude $\sim 150$ relative to the incoming intensity $\sim |F_0|^2$. This phenomenon is usually interpreted as a consequence of the spatial compression at the metal-vacuum interface of the waves impinging from the other side onto the dielectric-metal interface and tunneling through the thin metal layer.

For small enough frequencies ($\omega_0 \ll \omega_p$ = plasma frequency) the dispersion relation of the SPO approximately coincides with that of a free plane wave of light propagating in vacuum. If $\omega \approx \omega_0$, then the resonance condition $D_0 \ll 1$ at $\theta_0 = \theta_c$ for the determinant defined in Eq. (5) essentially equivalent to the dispersion relation $D_{sp} = 0$ of the SPO (see Eq. (6)), provided we make the replacement $k_{sp} \approx K_x = (\omega_0/c) \cdot n_1 \cdot \sin\theta_c \approx \omega_0/c$. In this case the parameters of the SPO approximately coincide with that of the forced oscillations, i.e. $a \approx A$ and $b \approx B$ (see the defining equations in Eqs. (5) and (6)). As is seen in Fig.3 (a), the reflected field is practically zero, and the incoming laser field is converted to surface plasmon oscillations in the ATR process. In the case of 100% conversion efficiency [7], we assume $|g_3| \approx |G_3| = g|F_0|$, where we have introduced the enhancement factor $g$ for the amplitudes. We have checked that at resonance the relative phase of $G_3$ and $F_0$ is zero, thus the SPO 'inherit' the phase of the incoming field, so we are allowed to take $g_3 \approx g \cdot F_0$.

## 3. Dressed state of a metallic electron embedded in the enhanced field of SPOs

In the present section the SPO field is considered as a ('strong') classical external field, and its effect in the Schrödinger equation of the electron will be taken into account non-perturbatively. In general, one can say that in the frame of this kind of description the transition probability amplitudes of high-order plasmon absorption or emission processes can be obtained in a one-step perturbation calculation, where the perturbation represents the effect of the other ('weak') agents (e.g., as below in the next section, in the case of light emission by the SPO, the interaction with the quantized mode of the spontaneously emerging photon will be considered as a perturbation).

First of all we have to have the explicit form of the electric field strength of the SPO, which is to be inserted into the interaction term in the Schrödinger equation. From Eqs. (1), (4), (6b), (6c), and on the basis of the considerations at the end of the previous section, in the metal layer the wave numbers and the electric field strength of the SPO can be expressed as

$$k_{2x} = k_{sp} = \frac{\omega_0}{c}\left[1 + \frac{1}{2(\varepsilon_R - 1)} + i\frac{\varepsilon_I}{2(\varepsilon_R - 1)^2}\right], \quad k_{2z} = \frac{\omega_0\sqrt{\varepsilon_R - 1}}{c}\left[1 + i\frac{\varepsilon_I}{2(\varepsilon_R - 1)}\right], \quad (7a)$$

$$\mathbf{E}_{sp}(\mathbf{r},t) = g|F_0|\exp(-x/2L_{sp})\exp(-z/2l_2)$$
$$\times[(\varepsilon_R - 1)^{-1}\mathbf{e_z}\cos(\omega_0 t'' + \varphi_0) - (\varepsilon_R - 1)^{-1/2}\mathbf{e_x}\sin(\omega_0 t'' + \varphi_0)], \quad (0 < z < d) \quad (7b)$$

$$L_{sp} \equiv |2\,\text{Im}[k_{sp}]|^{-1} \approx \frac{(\varepsilon_R - 1)^2}{2\pi\varepsilon_I}\lambda_0, \quad (7c)$$

$$l_2 \equiv \frac{\lambda_0}{4\pi(\varepsilon_R - 1)^{1/2}}, \qquad t'' \equiv t - (n_1 \sin\theta_c)\frac{x}{c} + \frac{\varepsilon_I}{2(\varepsilon_R - 1)^{1/2}}\frac{z}{c}. \qquad (7d)$$

The relative numerical accuracy of the above approximate expressions is of order of $\varepsilon_I / 2(\varepsilon_R - 1)^2$ (which is ~0.001 in our case). In Eq. (7c) we have introduced the the 'propagation length' $L_{sp}$ of the SPO (the $1/e$-length of the spatial energy distribution along the $x$-axis), which is $\sim 61.6 \times \lambda_0 = 49\mu m$ for $\varepsilon_R = 25.82$ and $\varepsilon_I = 1.63$. In the 'retarded time parameter' $t''$ introduced in Eq. (7d) the factor $n_1 \sin\theta_c \approx 1 + 1/[2(\varepsilon_R - 1)]$ is practically unity (the resonance angle $\theta_c$ differs from the angle of total reflection $\theta_t$ only by about one degree). In region 3 (in vacuum), with the same accuracy as above, we have

$$k_{3x} = k_{sp} = \frac{\omega_0}{c}\left[1 + \frac{1}{2(\varepsilon_R - 1)} + i\frac{\varepsilon_I}{2(\varepsilon_R - 1)^2}\right], \quad k_{3z} = \frac{\omega_0}{c\sqrt{\varepsilon_R - 1}}\left[1 + i\frac{\varepsilon_I}{2(\varepsilon_R - 1)}\right], \qquad (8a)$$

$$\mathbf{E}_{sp}(\mathbf{r},t) = g|F_0|\exp(-x/2L_{sp})\exp(z/2l_3) \\ \times [\mathbf{e_z}\cos(\omega_0 t' + \varphi_0) - (\varepsilon_R - 1)^{-1/2}\mathbf{e_x}\sin(\omega_0 t' + \varphi_0)]' \qquad (z < 0) \qquad (8b)$$

$$l_3 \equiv (\varepsilon_R - 1)^{1/2}\lambda_0/4\pi, \qquad t' \equiv t - (n_1 \sin\theta_c)\frac{x}{c} - \frac{\varepsilon_I}{2(\varepsilon_R - 1)^{3/2}}\frac{z}{c}. \qquad (8c)$$

Thus, at the metal-vacuum interface the incoming field given by Eq. (3) is transformed to SPO fields represented by Eqs. (7b) and (8b). Of course, this expressions are ment here only for positive values of $x$ (in the frame of our description the SPO is generated at $x = 0$, where the electron, by scattering on a potential irregularity, converts energy from the enhanced evanescent wave to the surface plasmon oscillations, securing the extra momentum needed for the resonant coupling). We note that, according to our approximate analytic estimate, the enhancement factor $g$ of the field amplitudes in Eqs. (7b) and (8b) behaves like $g \propto e^{(2\pi d/\lambda_0)\sqrt{\varepsilon_R + 1}}$ in the parameter range we are considering. We note that if we use the Drude free electron model, then we obtain $g \approx 30$ for system under discussion.

It is seen in Eq. (8b), that in vacuum the SPO has an elliptically polarized electric field whose longitudinal $x$-component is smaller by a factor of $(\varepsilon_R - 1)^{1/2} \approx 5$ than the $z$-component perpendicular to the metal surface. Inside the metal, on the other hand, the longitudinal component is larger by this same factor than the perpendicular one, so, this latter is smaller by a factor of $(\varepsilon_R - 1) \approx 25$ to compare with its value in vacuum. The fields inside the metal exponentially drop to their $1/e$-value within the distance of $\lambda_0/31.3 \approx 25.4 nm$. In the 'retarded time parameters' $t''$ and $t'$ defined in Eqs. (7c) and (8c), respectively, the factors of $z/c$ are much smaller than unity, so henceforth in the phases of the field we shall use the common notation $t' \approx t - (n_1 \sin\theta_c) x/c$, i.e. $\omega_0 t' = \omega_0 t - K_x x$. In the following we set $\varphi_0 = 0$ for the initial phase, which means that possible carrier-envelope phase difference effects [13] are not discussed here (in the experiments we have used $\sim 120\, fs$ 'long' pulses containing $\sim 46$ cycles).

The Schrödinger equation of an electron interacting with the SPO field in the $\mathbf{r}\cdot\mathbf{E}$-gauge

$$\hat{H}_e \psi_e(\mathbf{r},t) = i\hbar \partial_t \psi_e(\mathbf{r},t), \qquad \hat{H}_e = \frac{\hat{\mathbf{p}}^2}{2m} + V(z) + e\mathbf{r}\cdot\mathbf{E}_{sp}, \qquad (9)$$

where $\hat{\mathbf{p}} = -i\hbar\nabla$ is the electron's momentum operator in coordinate representation, and the strong external field $\mathbf{E}_{sp}$ has been given in Eqs. (7b) and (8b). In the present paper we use the Sommerfeld step potential model for the metallic electrons, thus in the Hamiltonian given by Eq. (9a) we take $V(z) = V_0[\Theta(-z) - 1]$, i.e. $V(z) = 0$ for $z < 0$ and $V(z) = -V_0$ for $z > 0$. This means that we confine ourselves to the discussion of processes taking place at the metal-vacuum interface. The stationary solutions of the unperturbed Schrödinger equation of an electron are superpositions of plane waves,

$$\psi_e(\mathbf{r},t) = \langle \mathbf{r}|\mathbf{p}\rangle_t \equiv (1/L)\exp[(i/\hbar)(p_x x + p_y y - Et)]\varphi_{E_z}(z), \qquad (10a)$$

where $E = E_x + E_y + E_z$ and $E_{x,y} = p_{x,y}^2/2m \geq 0$. The function $\varphi_{E_z}(z)$ is a solution of the well-known barrier problem of scattering on a step potential of depth $V_0 = E_F + A_w$ (see e.g. [42]), where $E_F$ is the Fermi energy ($= 5.51 eV$ for gold) and $A_w$ is the work function ($= 4.68 eV$ for gold), respectively. Depending on the energy associated to the $z$-motion, one has three kinds of solutions for the function $\varphi_{E_z}(z)$. We shall only need the form of $\varphi_{E_z}^{(0)}(z)$ for negative energies ($-V_0 \leq E_z < 0$),

$$\varphi_{E_z}^{(0)}(z) = \sqrt{\frac{m}{hp_0}} \begin{cases} e^{-\frac{i}{\hbar}p_0 z} + \dfrac{p_0 - iq_0}{p_0 + iq_0} e^{+\frac{i}{\hbar}p_0 z} & \text{for } z > 0 \\[1em] \dfrac{2p_0}{p_0 + iq_0} e^{+\frac{1}{\hbar}q_0 z} & \text{for } z < 0 \end{cases}, \qquad (10b)$$

$$p_0 \equiv \sqrt{2mV_0 - q_0^2}, \quad \frac{q_0^2}{2m} \equiv |E_z| = -E_z. \qquad (10c)$$

The electron propagates freely both inside and outside the metal (except in the negative $z$-direction for energies $E_z < 0$, when its momentum is imaginary), the interaction is limited to the $z = 0$ plane, where the reflection and transmission may take place. In both regions the fundamental solutions of Eq. (9) can be taken as modulated plane waves of the form

$$\overline{\psi}_e(\mathbf{r},t) = \langle \mathbf{r} - \mathbf{r}_{cl}(t') | \mathbf{p} \rangle_t = N \exp\{(i/\hbar)[\mathbf{p} \cdot (\mathbf{r} - \mathbf{r}_{cl}(t')) - Et]\} e^{-ih(t')}, \quad (11a)$$

$$\mathbf{r}_{cl}(t') \equiv \{x_{cl}(t'), 0, z_{cl}(t')\}, \quad x_{cl}(t') \equiv a g \mu_0 \lambdabar_0 \sin(\omega_0 t'), \quad z_{cl}(t') \equiv b g \mu_0 \lambdabar_0 \cos(\omega_0 t'), \quad (11b)$$

$$a = (\varepsilon_R - 1)^{-1/2}, \quad b = \begin{cases} (\varepsilon_R - 1)^{-1}, & 0 < z < d \\ 1, & z < 0 \end{cases}, \quad \mu_0 \equiv (eF_0/mc\omega_0), \quad \mu_0^2 = 10^{-18} I_0 \lambda_0^2, \quad (11c)$$

where $N$ is a normalization factor. The real function $h(t')$ in the exponent on the right hand side of Eq. (11a) is a simple periodic function of the retarded time $t'$, which drops out from the transition matrix elements to be calculated in the next section, so its explicit form is not interested here. In region 3 ($z < 0$) the $z$-component of the momentum of an evanescent electron is defined as

$$p_z = -\sqrt{2m(E_z - U_{sp})}, \quad U_{sp} \equiv g^2 \mu_0^2 (2mc^2)/8 = 5.14 \times 10^{-14} g^2 I_0 \lambda_0^2, \quad (z < 0) \quad (12)$$

where the 'ponderomotive energy shift' $U_{sp}$ of the electron comes as an average effect stemming from the SPO. This shift can play an important role in field enhanced multiphoton photoelectric emission from metal films [19-20]. $U_{sp}$ measured in $eV$ if on the right hand side of Eq. (12) $I_0$ and $\lambda_0$ measured in $W/cm^2$ and $\mu m$, respectively. In Eq. (11b) we have introduced the components of the classical trajectory $\mathbf{r}_{cl}(t)$, along which the electron oscillates in the SPO field according to the Newton equation $m\ddot{\mathbf{r}}_{cl} = -e\mathbf{E}_{sp}$. The dimensionless 'intensity parameter' $\mu_0$ introduced in Eq. (11b) plays a crucial role in nonlinear processes. Its numerical value can be calculated from the last equation in Eq. (11c), where the intensity $I_0$ and the central wavelength $\lambda_0$ of the incoming laser field measured in $W/cm^2$ and $\mu m$, respectively. It is important to note that in obtaining Eq. (11a) the $x$-dependence in $t'$ has been taken into account only parametrically, like in the very dipole approximation (where we use mere time-dependence). Similarly, we have considered the smooth profile functions of the amplitudes, $\exp(-x/2L_{sp})$ and $\exp(\mp z/2l_{2,3})$ merely as functions of the parameters $x$ and $z$. This is justified if the amplitudes of the classical oscillations $x_{cl}(t)$ and $z_{cl}(t)$ are much smaller than the wavelength $\lambda_0$, which can be satisfied if $g\mu_0 < 1$. Concerning the method of parametric substitution see [43-45], where it has also been shown that the de Broglie wavelength of the free electron has to be considerably smaller than the wave length of the radiation, for a relyable approximation. One can easily calculate that e.g.

$\lambda_{dB} / \lambda_0 = \hbar\omega_0 / pc \sim 10^{-4}$ even for thermal electrons. We note that the dressed states shown in Eq. (11a) are generalizations of the so-called Volkov states, which have been extensively used in the theory of strong-field laser-matter interaction phenomena from the very beginning [46]. An analogous but, on the other hand, a special form of these dressed states has been first used by Bunkin and Fedorov [47] in their treatment of laser-induced cold emission (optical tunneling of electrons) from metal surfaces.

## 4. Photon emission during multiple-plasmon scattering by a metallic electron

In order to describe spontaneous light emission by the dressed electron, we have to solve the Schrödinger equation of the joint system of an electron interacting with the SPO field and the quantized EM field of the photons,

$$i\hbar\partial_t |\Psi(t)\rangle = \hat{H}|\Psi(t)\rangle, \qquad \hat{H} = \frac{\hat{\mathbf{p}}^2}{2m} + V(z) + e\mathbf{r}\cdot\mathbf{E}_{sp} + e\mathbf{r}\cdot\hat{\mathbf{E}}_Q, \qquad (13)$$

where $\hat{\mathbf{p}} = -i\hbar\nabla$ is the electron's momentum operator, and the strong external field representing the surface plasmon oscillations, $\mathbf{E}_{sp}$ has been given in Eqs. (7b) and (8b). For the electric field strength $\hat{\mathbf{E}}_Q$ of the spontaneously emitted radiation in vacuum will be taken the following approximate form after Elson and Ritchie [34],

$$\hat{\mathbf{E}}_Q(\mathbf{r},t) = \sum_{\mathbf{k}} \int dk_z (4\hbar\omega/L^2)^{1/2}$$
$$\times [\hat{\mathbf{k}}(ck_z/\omega)\cos(k_z z + \eta) + i\mathbf{e}_\mathbf{z}(ck/\omega)\sin(k_z z + \eta)](\hat{a}^+_{\mathbf{k},k_z} e^{i\omega t} - \hat{a}_{-\mathbf{k},k_z} e^{-i\omega t})e^{-i\mathbf{k}\cdot\mathbf{r}}, \quad (z<0) \quad (14a)$$

$$\mathbf{k} \equiv (k_x, k_y, 0) \equiv (k\sin\theta)\hat{\mathbf{k}} = (k\sin\theta)(\cos\varphi, \sin\varphi, 0), \quad k_z = (\omega/c)\cos\theta, \qquad (14b)$$

$$\omega \equiv c\sqrt{k^2 + k_z^2}, \qquad \sin\eta = \frac{k_z \varepsilon_R}{\sqrt{(\varepsilon_R + 1)(k^2 + k_z^2 \varepsilon_R)}}, \qquad [\hat{a}_{\mathbf{k},k_z}, \hat{a}^+_{\mathbf{k}',k_z'}] = \delta_{\mathbf{k},\mathbf{k}'}\delta(k_z - k_z'). \qquad (14c)$$

(We show the expression only for the p-polarized components, because – as will be proved below – the emission of s-polarized photons are forbidden.) The summation and integration with respect to the wave numbers $\{\mathbf{k}, k_z\}$ of the emerging photons in Eq. (14a) is limited to $k$-space regions where $\omega < \omega_p$, because the metal layer is transparent for larger frequencies, and the mode functions have a different form to that given here. The form of the quantized field in Eq. (14a), given by Elson and Ritchie [31], corresponds to the approximation in which we consider a *semi-infinite lossless* metal ($\varepsilon_I = 0$) occupying the half-space $0 < z$. We have checked by a couple of numerical calculations that in the case of a *finite lossy* metal layer illustrated in Fig. 1, the reflection coefficients of the quantized modes are not considerably less than unity. In particular, if the polar angle $\theta$ (with respect to the *negative* $z$-axis) is close to $90°$, then the reflection coefficients are close to $100\%$, thus this approximation is justified (this is also supported *a posteriori* by the obtained results for the angular distributions of the spontaneously emitted photons, which is peaked around $\sim 70°$). We also note that for higher order harmonics ($n\omega > \omega_p$) the dielectric-metal system is practically transparent, and then, to a good approximation, we can work with the usual free-space expression of the electric field strength of the quantized radiation [35],

$$\hat{\mathbf{E}}(\mathbf{r},t) = i\sum_{\mathbf{k},s} \mathbf{e}_s(\mathbf{k})(2\pi\hbar\omega/L^3)^{1/2}[\hat{a}_{\mathbf{k},s}e^{i(\mathbf{k}\cdot\mathbf{r}-\omega t)} - \hat{a}^+_{\mathbf{k},s}e^{-i(\mathbf{k}\cdot\mathbf{r}-\omega t)}], \qquad (15)$$

where the subscript '$s$' refers to the two independent linear polarizations belonging to the propagation vector $\mathbf{k} = (k_x, k_y, k_z)$. Concerning the mode density, one essential difference between Eqs. (14a,b,c) and Eq. (15) can be immediately seen, namely the presence of the extra factors $ck_z/\omega = k_z/\sqrt{k^2 + k_z^2}$ and $ck/\omega = k/\sqrt{k^2 + k_z^2}$ on the right hand side of Eq. (14b). These factors are partly responsible for the modification of the kinematics (in other words, for the modification of the angular distribution) of the spontaneous emission, due to the presence of the metal boundary.

The transition matrix elements of the spontaneous emission can be calculated in a very similar manner as has been done in the earlier works of one of us [31-32]. For the initial and final states of the electron we take dressed states of the type given in Eq. (11a)

$$T_{fi} = -(i/\hbar)\int_{-\infty}^{\infty} dt \langle \overline{\psi}_{ef} | [\mathbf{e}\cdot\nabla V(z)] | \overline{\psi}_{ei} \rangle \exp[(i/\hbar)(E'-E)t]$$
$$\times (e/m\omega^2)(4\hbar\omega/L^3)^{1/2}(ck/\omega)\sin(kz+\eta)e^{i(\omega t - \mathbf{k}\cdot\mathbf{r})}, \qquad (16)$$
$$= -2\pi i \sum_n t_{fi}^{(n)} \delta(E - E' + n\hbar\omega_0 - \hbar\omega)$$

where the Dirac delta expresses the conservation of energy $E' + \hbar\omega = E + n\hbar\omega_0$. This means, that $n$ plasmons are absorbed by the electron and the energy is converted to the sum of the final electron energy and the energy of the spontaneously emitted photon. If the change in the electron's energy is small, then the spectrum of the emitted radiation is peaked at the harmonic frequencies $\omega = n\omega_0$. This is the case in *specular electron reflection* at the metal-vacuum interface. If we do not assume this specular reflection, then, in the frame of the present description, we cannot account for the narrow spectra measured in the experiments. The sum with respect to $n$ in Eq. (16) has been derived from the Fourier expansion of the periodic modulation factors in the dressed states by taking Eqs. (11a,b) into account, and using the Jacobi-Anger formula $\exp(iz\sin\varphi) = \sum_{n=-\infty}^{\infty} J_n(z) e^{in\cdot\varphi}$, where $J_n(z)$ denotes the ordinary Bessel function of first kind of order $n$ [48]. From Eq. (16) we immediately see that *the emission of s-polarized photons are forbidden* (the gradient of $V(z)$ is necessarily pointing to the $z$-direction, so a vector **e** in the plane of the interface is of course orthogonal to that). On the basis of Eq. (16) the production rates of the $n^{th}$ harmonic photons can be brought to the form

$$\frac{dP_n}{d\Omega} = \nu_0 \alpha f^2(\theta) G^2(\upsilon)$$

$$\times \left| \sum_k n F(u_{n+k}) J_{n+k}[ag\mu_0(c/\hbar\omega_0)(p_x - p'_x)] F(u_{n+k}) I_k[g\mu_0(c/\hbar\omega_0)(|p_z| + |p'_z|)] \right|^2, \quad (17)$$

$$f(\theta) \equiv \frac{2\varepsilon_R(\sin 2\theta)}{\sqrt{(\varepsilon_R + 1)(\sin^2\theta + \varepsilon_R \cos^2\theta)}}, \quad (17a)$$

$$F(u_{n+k}) \equiv \frac{1}{1 + iu_{n+k}}, \quad u_{n+k} \equiv 2[p_x - p'_x + (n+k)\hbar K_x - \hbar k_x] L_{sp}/(n+k)\hbar, \quad (17b)$$

$$G(\upsilon) \equiv \frac{\sin\upsilon}{\upsilon}, \quad \upsilon \equiv (p_y - p'_y - \hbar k_y)L/\hbar, \quad (17c)$$

where $\alpha = e^2/\hbar c \cong 1/137$ is the fine structure constant, and $\nu_0 = \omega_0/2\pi$. In obtaining Eq. (17) we have used the relation $J_n(iz) = i^n I_n(z)$, which connects the ordinary Bessel functions $J_n(z)$ and the modified Bessel functions $I_n(z)$ of the first kind of integer order. In order to complete the calculation, the above formula for the production rate still should be integrated with respect to the possible initial and final electron momenta **p** and **p**′, respectively, by taking into account the conservation of energy $E' + \hbar\omega = E + n\hbar\omega_0$ and the corresponding Fermi distribution. This has been approximately done in a similar case in [31], with the conclusion that this procedure leads to a spreading in energy of order of $\Delta E \approx 4 \times kT \approx 0.1 eV$ (at room temperature, $T = 293K$, $kT \approx 0.025 eV$). The relative energy spread $\Delta E / E_F \approx 0.018$ would correspond to the spectral width $\Delta\lambda \approx 0.018 \times \lambda_0 = 14.3 nm$, which is quite close to the experimental value [38]. In Eq. (17) $G^2(\upsilon)$ vanishes for $L \to \infty$, unless $k_y$ is zero. Thus, the wave vector of the emitted photon lies in the $x-z$ plane (the scattering plane). In Eq. (17) the angle between the negative $z$-axis and the k-vector of the emitted photon is denoted by $\theta$, thus its wave vector has the form $\vec{k} = (\omega/c)(\sin\theta, 0, \cos\theta)$.

In the rest of the present section we give some illustrations of the numerical results on the basis of the formula for the differential production rates given by Eq. (17).

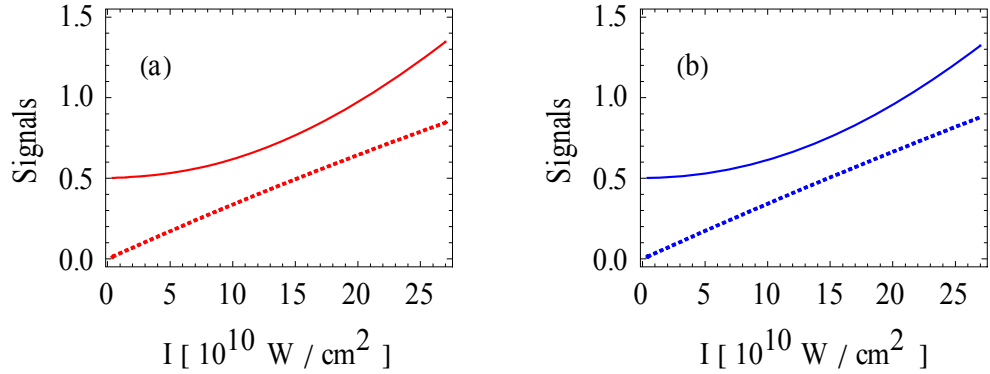

**Figure 4.** Shows, the intensity dependence of the fundamental (a) and the second harmonic (b) dimensionless production rates $(dP_n/d\Omega)/(4\times 10^{-6}\alpha v)$ according to the formula given by Eq. (17). The dashed and the full lines refer to results with no field enhancement and with enhancement of magnitude $g^2 = 150$. (Because the enhanced signals are much larger than the signals with no enhancement, we have divided their value by 1500 for the fundamental and by 1450 for the second harmonics, respectively.

The dashed lines in Fig.4 show that in the intensity range we have considered the linear dependence is reproduced, which comes out when one treats the SPO field pertubatively. The strong deviation from the pertubative results are clearly seen. It is remarkable that the enhanced signals do not saturate, at least in this intensity range, rather, their slope remains positive for large intensities. This behaviour is in agreement with our measurements [38].

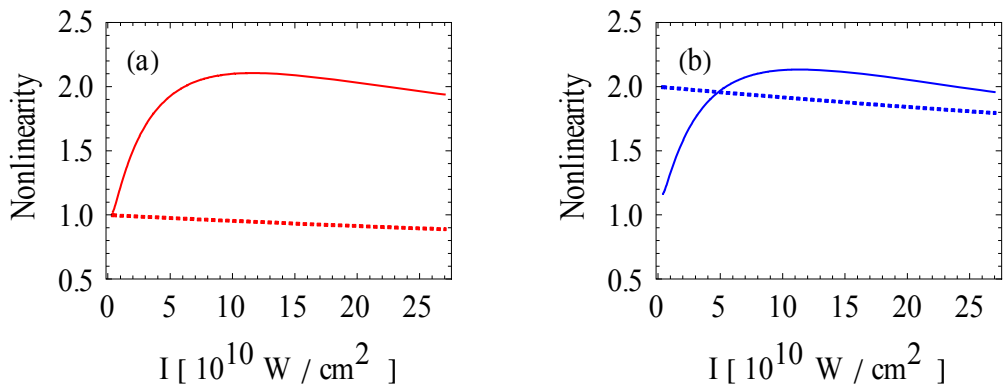

**Figure 5.** In this figure we show the dependence of the "degree of nonlinearity", $\partial\log S/\partial\log I$, where $S$ stands for the signal intensity. The labeling of the curves are the same as in Figure 4. In case of no enhancement the slope of the signals follow the pertubative dependence. The dased lines show that (a): $\partial\log S_1/\partial\log I \cong 1$ and (b): $\partial\log S_2/\partial\log I \cong 2$ in the whole intensity range.

As is shown in Fig. 5, without enhancement, the well-known pertubative resuls are reproduced, i.e. $S_1 \sim I$ for the fundamental, and $S_2 \sim I^2$ for the second harmonic radiation. On the other hand, the degree of nonlinearly of the signals stemming from the enhanced field, strongly deviates from the pertubative results in a complicated way.

It can be shown that for low intensities of the incoming radiation (when e.g. $\mu_0 = 10^{-4}$ for a Ti: Sapphire laser of intensity $I_0 = 10^{10} W/cm^2$), even by taking into account the field enhancement, the angular distribution of the emitted light is essentially governed by the function $f(\theta)$ defined in Eq. (17a). This function does not depend on the intensity, but it is sensitive to the frequency through the frequency-dependent permittivity $\varepsilon_2(\omega)$. In Fig. 6 we have plotted the angular distributions of the fundamental ($\varepsilon_R = 25.82$) and of the second harmonic ($\varepsilon_R = 5.71$) signals, by using exlusively the function $f(\theta)$. According to this Fig. 6, the distribution of the fundamental radiation is peaked around $\theta_{max}^{(1)} = 70°$, and the second harmonic has a maximum at about $\theta_{max}^{(2)} = 60°$, and this latter is broader than the former one. For the detailed discussion of the angular distribution for larger incoming intensities than $\sim 10^{10} W/cm^2$ we have to thoroughly investigate the numerical behaviour of the sum of the products of the Bessel functions in the formula of the production rates given by Eq. (17). This will be one of the subjects of our future work concerning non-linear surface plasmonics.

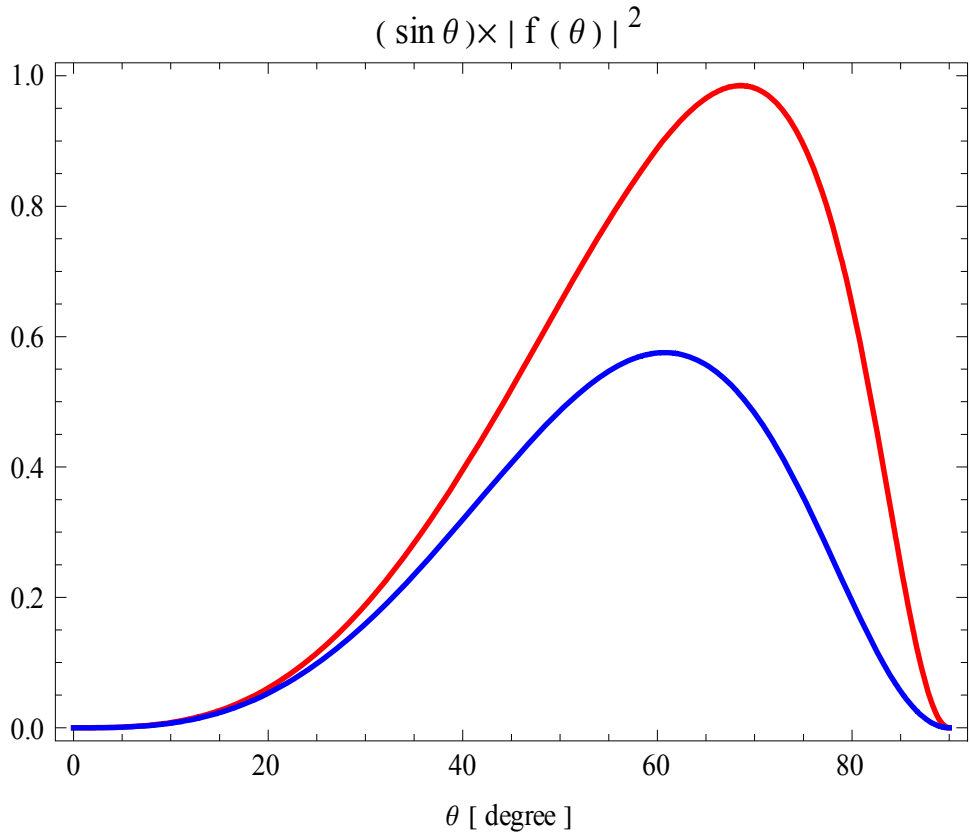

**Figure 6.** Shows the angular dependence of the fundamental (red curve) and the second harmonic (blue curve) normalized photon production rates in the low-intensity case, when $I_0 = 10^{10} W/cm^2$ has been assumed, and the enhancement factor has been taken to be $g = \sqrt{150}$. In the figure the function $1.6 \times (\sin\theta)|f(\theta)|^2$ has been plotted on the basis of the defining equation Eq. (17a). For the quantitative difference between the curves the use of the different dielectric constants is responsible. For $\omega = \omega_0$ the value $\varepsilon_R = 25.82$, and for $\omega = 2\omega_0$ the value $\varepsilon_R = 5.71$ has been taken.

## 5. Summary

In the present paper we have given a theoretical analysis of the enhanced surface-plasmon-assisted spontaneous radiation of a metallic electron. First, in Section 2 we have discussed the kinematics of the generation of the occasionally very high-intensity evanescent electromagnetic fields and surface plasmon oscillations at the metal-vacuum interface of a dielectric-metal-vacuum system. We have determined the spatial distribution and the amplitude of this field, and have given a couple of numerical examples which are relevant in our recent experiments. In Section 3, after having determined the interaction term to be put into the Schödinger equation, the dressed quantum states of the electrons embedded into the high-intensity SPO field have been derived. In Section 4, on the basis of these dressed states, we have determined in a one-step perturbation calculation the transition matrix elements of the multiple-plasmon-assisted spontaneous emission of photons in the SPO environment. Here we have given a general formula for the absolute production rates of of the harmonic components for arbitrary high order. As a first result, it follows from our theory that the emitted radiation must exclusively be p-polarized, in agreement with the experiments. We have made an estimate on the relative bandwidth of the radiation, and illustrated the unusual characteristics of the emitted radiation with the help of some typical numerical examples, as applications of our analytic results. It has turned out that, due to the large enhancement of the exciting laser fields in the vicinity of the interfaces, a considerable distortion of the intensity distribution of the signal appears, like in our recent experiments. Finally, we have given a special example, relevant only in the low-intensity case, for the angular distribution of the emitted fundamental and second harmonic radiation. In general, we have found a good qualitative agreement between the predictions of the present theory and our recent experimental findings on enhanced and nonlinear light emission at metal surfaces. According to the appeerent harmony between theory and experiment, we conjecture that the light signals measured in our recent experiment are stemming from multiple SPO conversion during the scattering on the free electrons which are reflected specularly at the metal surface.

We think that the theoretical method presented here may also find applications in analysing other plasmon-induced nonlinear processes, like for instance high-order electron emission from thin metal films.

**Acknowledgements.** This work has been supported by the Hungarian National Scientific Research Foundation OTKA, Grant No. K73728. The authors thank the unknown referees for the constructive critics and for many valuable suggestions made by them, most of which have been taken into account in the final version of the present paper.

**References**

[1] J. A. Stratton, *Electromagnetic theory* (New Jersey : Wiley, 2007)

[2] A. Sommerfeld, *Ann. der Phys.* **28,** 665 (1909)

[3] E. Burstein and F. de Martini (Eds.), *Polaritons* (London: Pergamon, 1974)

[4] A. A. Maradudin, V. M. Agranovich and D. L. Mills (Eds.), *Surface Polaritons* (Amsterdam: North-Holland, 1982)

[5] H. Raether, *Excitation of plasmons and interband transitions by electrons* (Berlin: Springer, 1980), H. Raether H, *Surface plasmons* (Berlin: Springer, 1988)

[6] A. V. Zayats, I. I. Smolyaninov , *J. Opt. A: Pure Appl. Opt.* **5**, S16 (2003)

[7] A. V. Zayats, I. I. Smolyaninov  and A. A. Maradudin , *Physics Reports* **408,** 131 (2005)

[8] M. I. Stockman, S. V. Faleev and D. J. Bergman, *Phys. Rev. Lett*. **87**, 167401 (2001)

[9] E. Altewischer, M. P. van Exter and J. P. Woerdman, *Nature* **418**, 304 (2002)

[10] W. L. Barnes, A. Dereux and T. W. Ebbesen, *Nature* **424**, 824 (2003)

[11] S. E. Irvine and A. Y. Elezzabi, *Appl. Phys. Lett*. **86**, 264102 (2005)

[12] S. Fasel, M. Halder, N. Gisin and H. Zbinden, *New J. Phys.* **8**, 13 (2006)

[13] S. E. Irvine, P. Dombi, Gy. Farkas and A. Y. Elezzabi, *Phys. Rev. Lett*. **97**, 146801 (2006)

[14] K. J. Chau, M. Johnson and A. Y. Elezzabi, *Phys. Rev. Lett*. **98**, 133901 (2007)

[15] M. I. Stockman, M. F. Kling, U. Kleineberg and F. Krausz, *Nature Photon* **1**, 539 (2007)

[16] M. Bumki, E. Ostby, V. Sorger, E. Ulin-Avila, L. Yang, X. Zhang and K. Vahala, *Nature* **457**, 455 (2009)

[17] M. L. Brongersma  and P. G. Kirk (Eds.), *Surface plasmon nanophotonics* (Dordrecht: Springer, 2007)

[18] U. Fano, *Ann.  der Phys.* **32,** 393 (1938)

[19] T. Tsang, T. Srinivasan-Rao and J. Fischer, *Phys. Rev. B* **43**, 8870 (1991)

[20] M. Aeschlimann, C. A. Schmuttenmaer, H. E. Elsayed-Ali, R. J. D. Miller, J. Cao and D. A. Mantell, *J. Chem. Phys*. **102**, 8606 (1995)

[21] M. R. Roth, C. N. Panoiu, M. M. Adams and M. R. Jr. Osgood, *Opt. Express* **14**, 2921 (2006)

[22] S. Kim, J. Jin, Y-J. Kim, I-Y. Park, Y. Kim and S-W. Kim, *Nature* **453**, 757 (2008)

[23] J. Wu, H. Qi and H. Zeng, *Phys. Rev. A* **77**, 053412 (2008)

[24] P. Lu, J. Wu, H. Qi and H. Zeng, *Appl. Phys. Lett.* **93**, 201108 (2008)

[25] J. Wu, H. Qi and H. Zeng, *Appl. Phys. Lett.* **93**, 051103 (2008)

[26] A. L'Huillier, K. J. Schaefer and K. Kulander, *J. Phys. B. At. Mol. Opt. Phys.* **24**, 3315 (1991)

[27] M. Lewenstein, P. Balcou, M. Yu. Ivanov, A. L'Huillier and P. B. Corkum, *Phys. Rev. A* **49**, 2117 (1994)

[28] F. Krausz and M. Ivanov, *Rev. Mod. Phys.* **81**, 163 (2009)

[29] W. L. Barnes, *J. Mod. Opt.* **45,** 661 (1998)

[30] D. von der Linde, T. Engers, G. Jenke, P. Agostini, G. Grillon, E. Nibbering, A. Mysyrowicz and A. Antonetti, *Phys. Rev. A*, **52**, 25 (1995)

[31] S. Varró and F. Ehlotzky, *Phys. Rev. A*, **49**, 3106 (1994)

[32] S. Varró and F. Ehlotzky, *Phys. Rev. A*, **54**, 3245 (1996)

[33] E. Kretschmann and H. Raether, *Zeitschr. für Naturforsch.* **23a,** 2135 (1968)

[34] J. M. Elson and R. H. Ritchie, *Phys. Rev. B* **4**, 4129 (1971)

[35] J. Bergou J and S. Varró, *J. Phys. A: Math. Gen*. **14**, 2281 (1981)

[36] N. Kroó, S. Varró, Gy. Farkas, P. Dombi, D. Oszetzky, A. Nagy and A. Czitrovszky, *J. Mod. Opt.* **54,** 2679 (2007)

[37] N. Kroó, S. Varró, Gy. Farkas, P. Dombi, D. Oszetzky, A. Nagy and A. Czitrovszky, *J. Mod. Opt.* **55,** 3203 (2008)

[38] N. Kroó, Gy. Farkas, P. Dombi and S. Varró, *Opt. Express* **16**, 21656 (2008)

[39] E. N. Economou, *Phys. Rev.* **182,** 539 (1969)

[40] J. J. Burke, G. I. Stegeman and T. Tamir, *Phys. Rev. B* **33**, 5186 (1986)

[41] P. B. Johnson and R. W. Christy, *Phys. Rev. B* **6**, 4370 (1972)

[42] D. Bohm, *Quantum theory*. Chapter 11 (Englewood Cliffs: Prentice-Hall, 1957)

[43] J. Bergou and S. Varró, *J. Phys. A: Math. Gen.* **13,** 3553 (1981)

[44] K. Drühl and J. K. McIver, *J. Math. Phys.* **24,** 705 (1983)

[45] H. Reiss, *Phys. Rev. A* **63**, 705 (2000)

[46] N. Kylstra, C. J. Joachain and M. Dörr, pp 15-36. In D. Batani, C. J. Joachain, S. Martellucci and A. N. Chester (Eds.), *Atoms, Solids and Plasmas in Super-intense Laser Fields* (New York: Kluwer, 2001)

[47] F. V. Bunkin and M. V. Fedorov, *Sov.Phys. JETP* **21**, 896 (1965)

[48] Erdélyi (Editor), *Higher transcendental functions. Vol. II*. Formula 7.2.4(26) (New York: McGraw-Hill, 1953)

[49] M. Born and E. Wolf, *Principles of optics*, Section 13.3 (London: Pergamon, 1959)